\documentclass[a4paper,fleqn,usenatbib,letter]{mnras}

\usepackage{graphicx}	
\usepackage{amsmath}	
\usepackage{amssymb}	

\title[A shock front at the radio relic of A2744]{A shock front at the radio relic of Abell 2744}

\author[D. Eckert et al.]
{D. Eckert$^{1}$\thanks{E-mail:
Dominique.Eckert@unige.ch}, M. Jauzac$^{2,3}$, F. Vazza$^{4}$, M. S. Owers$^{5,6}$, J.-P. Kneib$^{7}$, C. Tchernin$^{8,1}$,
\newauthor
H. Intema$^{9,10}$ , K. Knowles$^{3}$
\\
\\
$^{1}$Astronomy Department, University of Geneva, 16 ch. d'Ecogia, CH-1290 Versoix, Switzerland\\
$^{2}$Centre for Extragalactic Astronomy, Department of Physics, Durham University, Durham DH1 3LE, U.K.\\
$^{3}$Astrophysics and Cosmology Research Unit, School of Mathematical Sciences, University of KwaZulu-Natal, Durban 4041, South Africa\\
$^{4}$Hamburg Observatory, Gojansbergsweg  112, 21029 Hamburg, Germany\\
$^{5}$Australian Astronomical Observatory, PO Box 915, North Ryde, NSW 1670, Australia\\
$^{6}$Department of Physics and Astronomy, Macquarie University, NSW, 2109, Australia\\
$^{7}$Laboratoire d'Astrophysique, Ecole Polytechnique F\'ed\'erale de Lausanne (EPFL), Observatoire de Sauverny, CH-1290 Versoix, Switzerland\\
$^{8}$Center for Astronomy, Institute for Theoretical Astrophysics, Heidelberg University, Philosophenweg 12, 69120 Heidelberg, Germany\\
$^{9}$Leiden Observatory, Leiden University, P.O. Box 9513, NL-2300 RA Leiden, The Netherlands\\
$^{10}$National Radio Astronomy Observatory, 1003 Lopezville Road, Socorro, NM 87801-0387, USA
}

\date{Accepted }

\pubyear{2016}

\begin{document}
\label{firstpage}
\pagerange{\pageref{firstpage}--\pageref{lastpage}}
\maketitle

\begin{abstract}
Radio relics are Mpc-scale diffuse radio sources at the peripheries of galaxy clusters which are thought to trace outgoing merger shocks. We present \emph{XMM-Newton} and \emph{Suzaku} observations of the galaxy cluster Abell 2744 ($z=0.306$), which reveal the presence of a shock front 1.5 Mpc East of the cluster core. The surface-brightness jump coincides with the position of a known radio relic. Although the surface-brightness jump indicates a weak shock with a Mach number $\mathcal{M}=1.7_{-0.3}^{+0.5}$, the plasma in the post-shock region has been heated to a very high temperature ($\sim13$ keV) by the passage of the shock wave. The low acceleration efficiency expected from such a weak shock suggests that mildly relativistic electrons have been re-accelerated by the passage of the shock front.
\end{abstract}

\begin{keywords}
galaxies: clusters: individual: A2744 -- X-rays: galaxies: clusters
\end{keywords}

\section{Introduction}

In the standard hierarchical structure formation scenario, galaxy clusters are expected to form through subsequent mergers of smaller entities. In this process, the plasma contained within the merging subclusters collides, launching Mpc-scale shock waves with Mach numbers in the range 1.5-3 which contribute to the thermalization of the plasma \citep[e.g.][]{miniati01,ryu03}. This picture has received increasing observational support lately thanks to the discovery of shock fronts in a handful of merging clusters \citep{markevitch02,markevitch05,owers11,owers14,russell10}. Additionally, merger shocks are also expected to (re-)accelerate particles through diffusive shock acceleration \citep[DSA, e.g.][]{ensslin98,hoeft08,kang13}. This phenomenon is thought to give rise to  \emph{radio relics}, i.e. irregular, diffuse, steep-spectrum radio sources at the periphery of merging clusters \citep[e.g.][]{rottgering97,vanweeren10,bonafede12}. 

While in a few cases a shock front has been found coincident with a radio relic \citep{finoguenov10,ogrean13b}, X-ray observations tend to prefer lower Mach numbers compared to the values expected from DSA based on the radio spectrum \citep{akamatsu13,ogrean13,vanw16}. Some X-ray-detected shock fronts are not associated with radio emission \citep[Abell 2146,][]{russell11}, and the acceleration efficiency of electrons and protons differs from that expected from DSA \citep{vazza14,vazza15}. An alternative possibility is that radio relics arise from the re-acceleration of pre-existing cosmic-ray electrons by turbulence induced by the passage of the shock front \citep{fujita16}. Therefore, while there is usually a connection between shock fronts and radio relics, different systems provide contradicting results and the acceleration mechanism is still poorly understood. 

Abell 2744 (hereafter A2744) is a massive cluster \citep[$M_{200}\sim2\times10^{15}M_\odot$,][]{medezinski16} at a redshift of 0.306 \citep{owers11}. It is located at the crossroads of several filaments \citep{eckert15} and it is experiencing a merger of at least four individual entities. Radio observations of this system have reported the presence of a central radio halo and also of a peripheral radio relic of size $1.6\times0.3$ Mpc located roughly 1.5 Mpc East of the cluster core \citep{govoni01}, which may indicate the presence of a shock front in this region \citep{orru07}. Recently, \citet{ibaraki14} reported a very high temperature of $\sim13$ keV in the relic region, which may be suggestive of shock heating.  

In this paper, we report the discovery of a shock front associated with the radio relic of A2744. The paper is organized as follows. In Sect. \ref{sec:data} we describe the available X-ray data and the analysis procedure. Our results are presented in Sect. \ref{sec:results} and discussed in Sect. \ref{sec:disc}. At the redshift of A2744, the standard $\Lambda$CDM cosmology corresponds to $1^{\prime\prime}=4.5$ kpc. The uncertainties are given at the $1\sigma$ level.

\section{Data}
\label{sec:data}

\subsection{XMM-Newton}
A2744 was observed by \emph{XMM-Newton} on December 18-20, 2014 for a total exposure of 110 ks (OBSID 074385010, PI: Kneib). We analyzed the data using the XMMSAS software package v13.5 and the ESAS data analysis scheme \citep{snowden08}. After cleaning the event files by excluding the time periods affected by soft proton flares, we obtain 96 ks (MOS1), 97 ks (MOS2), and 87 ks (PN) of usable data. We extracted a count image in the [0.5-1.2] keV band and computed the corresponding exposure map to correct for vignetting effects. A model for the particle background was created from a large collection of filter-wheel-closed observations which were then rescaled to match the particle background count rate in the A2744 observation. This dataset was presented in \citet{eckert15} and Jauzac et al. (subm.). For more details on the analysis procedure, we refer the reader to these papers.

\subsection{Suzaku}
The NE field of A2744 around the radio relic was targeted by \emph{Suzaku} on November 20-22, 2013 for a total of 70 ks (OBSID 808008010). A three-pointing mosaic of the outer regions of A2744 also exists in the archive and was presented by \citet{ibaraki14}. For the purpose of this study we only focus on the NE pointing. We analyzed this observation using the \emph{Suzaku} FTOOLS v6.17 and the corresponding calibration database. We reprocessed the data using the \texttt{aepipeline} tool, filtering out the time periods when the geomagnetic cut-off rigidity was $<6$ GeV. We extracted spectra from the three XIS chips in the relic region and in a background region located 12.5 arcmin NE of the cluster core, where no cluster emission is detected. We used the \texttt{xisnxbgen} tool to extract model particle background spectra from dark-Earth data. The particle background spectra were fit using a phenomenological model and added to the global spectral model to preserve the statistical properties of the data, following the approach presented in \citet{degrandi16}. 

\section{Results}
\label{sec:results}

\subsection{Surface-brightness profile}

\begin{figure*}
\hbox{\includegraphics[width=0.5\textwidth]{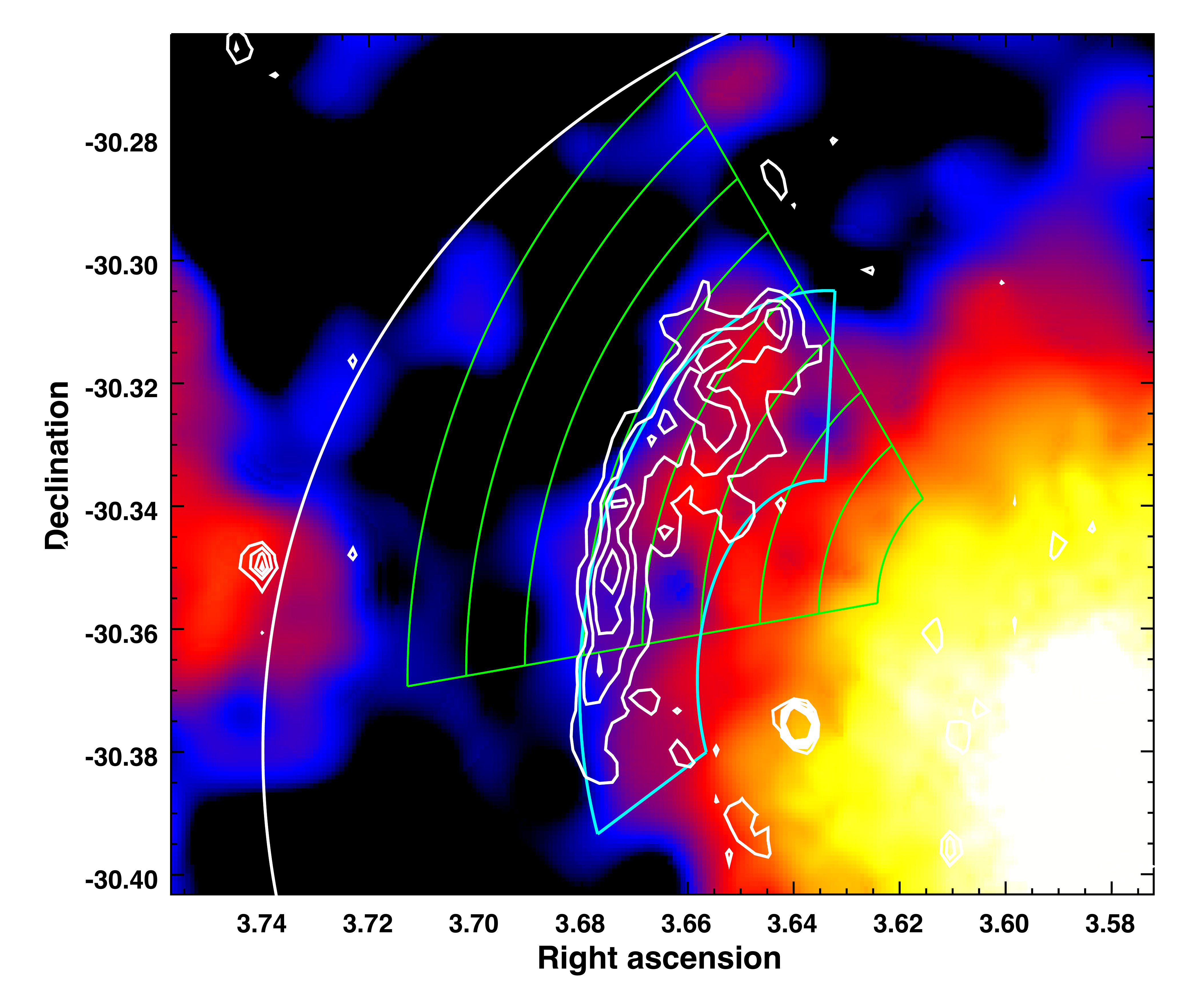}\includegraphics[width=0.5\textwidth]{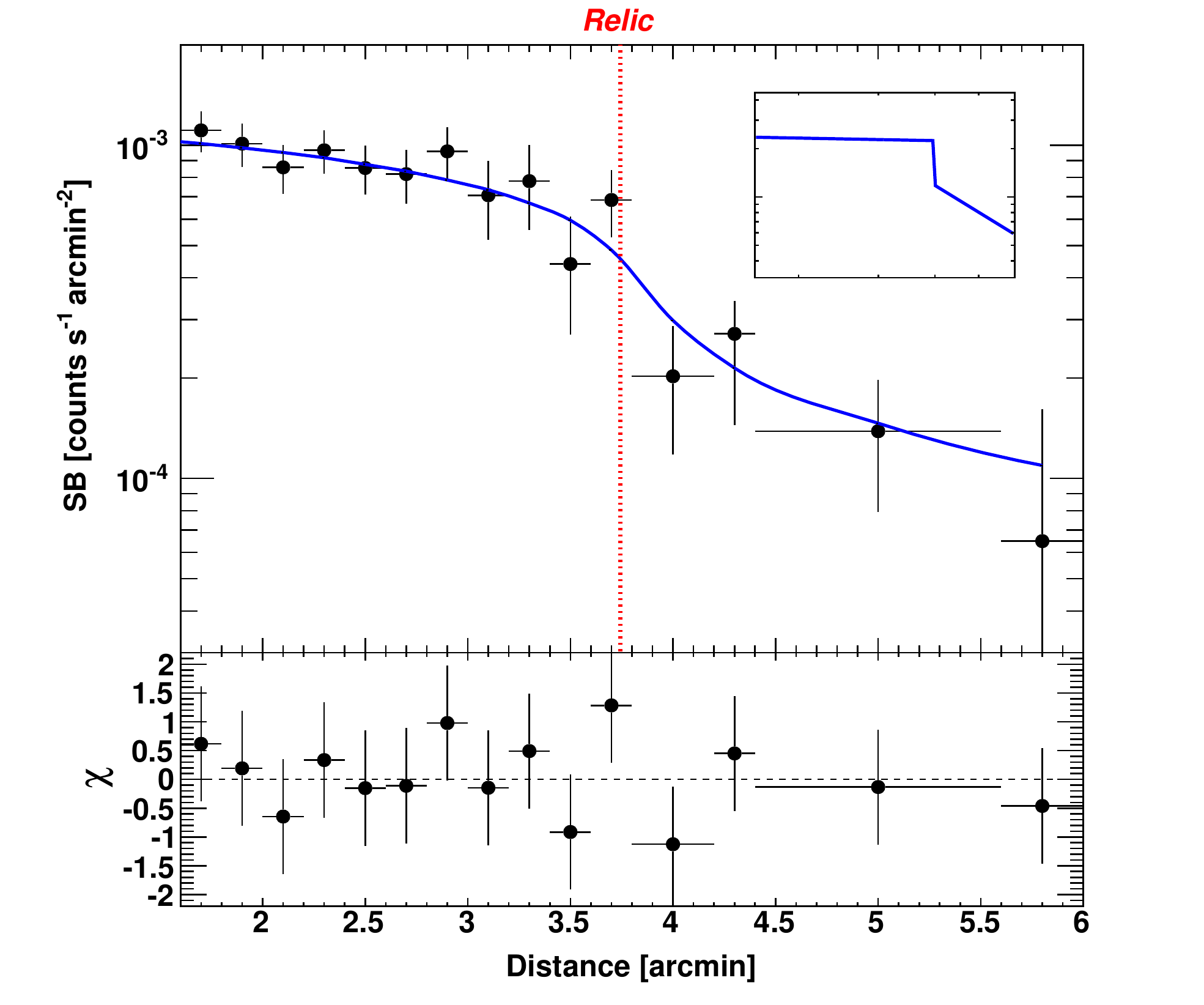}}
    \caption{\emph{Left:} Adaptively-smoothed \emph{XMM-Newton}/EPIC image of A2744 in the [0.5-1.2] keV band around the radio relic. The white contours show the radio emission at 330 MHz observed by GMRT. The contours are at 250, 500, 800, 1100, 1500 $\mu$Jy/beam. The significance of the lowest contour is 3 sigma. The green sector is the region used to extract the surface-brightness profile. The cyan sector displays the regions used to extract the \emph{XMM-Newton} spectrum. The approximate location of $R_{200}$ for A2744 is shown by the white circle. \emph{Right:} EPIC surface-brightness profile across the radio relic in the [0.5-1.2] keV band, fit with a broken power law model projected along the line of sight and convolved with the PSF of \emph{XMM-Newton}. The corresponding 3D gas density model is shown in the inset in the top-right corner. The bottom panel displays the residuals from the fit. The dashed red line shows the location of the eastern edge of the radio relic.}
    \label{fig:jump}
\end{figure*}

In the left-hand panel of Fig. \ref{fig:jump} we show the co-added \emph{XMM-Newton}/EPIC image of A2744 in the [0.5-1.2] keV band centered on the radio relic. The image was corrected for vignetting and the model particle background was subtracted. The resulting image was then adaptively smoothed using the XMMSAS task \texttt{asmooth}. For comparison, we also show GMRT radio contours at 330 MHz. The GMRT image has a restoring beam size of $13.4^{\prime\prime}\times8.0^{\prime\prime}$ and an RMS background noise of 100 $\mu$Jy per beam (Mulcahy et al. in preparation). A sharp drop in X-ray surface brightness can be observed beyond the eastern edge of the relic. 

To confirm this statement, we used \texttt{Proffit} v.1.3 \citep{eckert11} to measure the brightness profile across the relic in the sector displayed in green in Fig. \ref{fig:jump}\footnote{Note that the binning used for the surface brightness profile does not correspond exactly to the sectors shown in the figure.}. The resulting surface-brightness profile is shown in the right-hand panel of Fig. \ref{fig:jump}. A flat profile is observed inside the relic, followed by a sharp decline in the measured surface brightness. This behaviour is characteristic of a density discontinuity. Thus, we fitted the surface-brightness profile with a broken power-law model of the form 

\begin{equation}\rho(r)=\left\{\begin{array}{ll} \mathcal{C}~r^{-\Gamma_{\rm in}}, & \mbox{ if }r<r_{\rm break}\\
\mathcal{C}\frac{n_{\rm out}}{n_{\rm in}}~r^{-\Gamma_{\rm out}}, & \mbox{ otherwise}\end{array}\right.\end{equation}

\noindent The model thus has 5 free parameters ($\mathcal{C}$, $\Gamma_{\rm in}$, $\Gamma_{\rm out}$, $r_{\rm break}$, and $n_{\rm in}/n_{\rm out}$). This model is projected along the line of sight and convolved with the PSF of \emph{XMM-Newton}. The best-fit model is displayed in the right-hand panel of Fig. \ref{fig:jump}, confirming the presence of a brightness jump coincident with the eastern edge of the radio relic. The model provides an accurate description of the data ($\chi^2=4.7$ for 10 d.o.f.) and yields a significant improvement compared to a single power-law fit ($\chi^2=26.9$ for 13 d.o.f.), which according to the F-test corresponds to a null-hypothesis probability of $4\times10^{-4}$ ($3.6\sigma$ level). The best-fit model returns density slopes $\Gamma_{\rm in}=0.1\pm0.3$ and $\Gamma_{\rm out}=1.7\pm0.7$ inside and outside the relic, respectively. The outer slope of the gas density profile is consistent with the value of $-2$ usually observed in cluster outskirts \citep[e.g.][]{e12,morandi15}. The deprojected density jump at the front is $n_{\rm in}/n_{\rm out}=1.9_{-0.4}^{+0.6}$. This analysis demonstrates that a density discontinuity is present, coincident with the eastern edge of the radio relic. Note that the exact value of the density jump obviously depends on the choice of the extraction sector for the surface-brightness profile. Thus, in case the discontinuity does not follow exactly the position of the relic, our measurement of the density jump may be under-estimated.

\subsection{Spectral properties}

We extracted spectra in the relic region from both \emph{XMM-Newton} and \emph{Suzaku} data to investigate the thermodynamic properties of the plasma inside the density discontinuity. The corresponding region is shown in the left-hand panel of Fig. \ref{fig:jump}. In both cases, we used offset regions to estimate the background parameters. For the exact definition of the offset regions we refer to Extended Data Fig. 2 of \citet{eckert15}. To model the background spectrum, we followed the approach presented in \citet{eckert14}. The background spectrum was fit with a 4-component model including a phenomenological model for the particle background, a power law for the cosmic X-ray background and two APEC thin-plasma models for the Galactic halo and the local hot bubble. Spectral fitting was performed using XSPEC v12.9 and the C-statistic estimator. The background spectrum was extracted from four different offset regions in the field and no variation in the background parameters was found across the field of view \citep[see][for a detailed description of the background model in the region of A2744]{eckert15}. To model the spectrum of the relic region, the normalization of the sky background parameters was renormalized by the ratio of the area of the source region to that of the background region. Conversely, the normalization of the particle background model was left free while fitting. The source itself was modeled with a single-temperature APEC model absorbed by the Galactic $N_H$ \citep[$1.5\times10^{20}$ cm$^{-2}$,][]{kalberla}. The metal abundance was fixed to the canonical value of $0.25Z_\odot$ \citep{lm08b}. 

\begin{figure}
\includegraphics[width=\columnwidth]{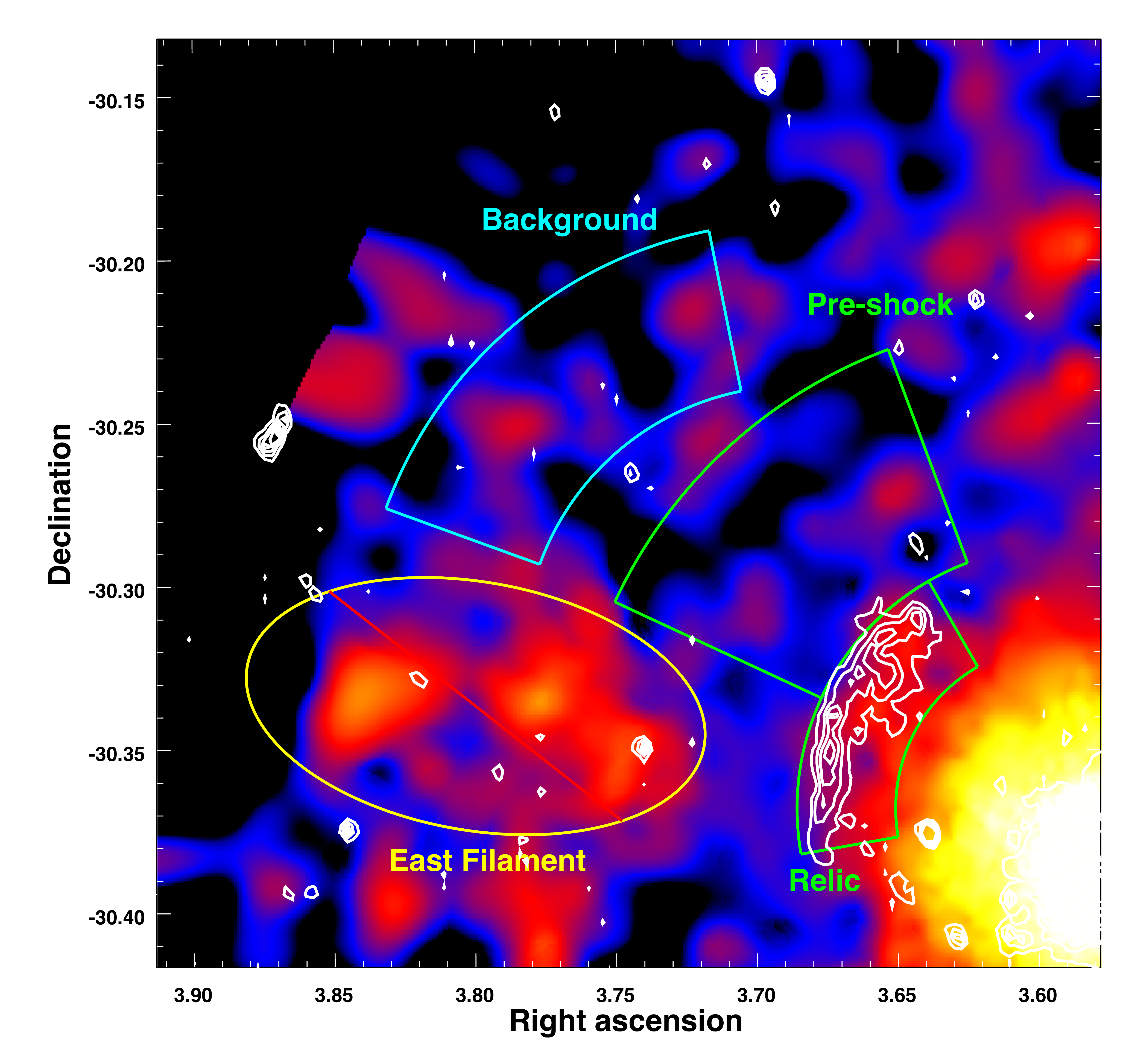}
    \caption{Same as Fig. \ref{fig:jump} (left), but showing the regions defined for the spectral analysis of the \emph{Suzaku} data. The green regions show the location of the relic and pre-shock regions, whereas the cyan region shows the region used to extract the background spectrum. The yellow ellipse denotes the approximate limits of the Eastern filament of A2744 \citep{eckert15}, which was masked during the analysis to avoid biasing the results (red line).}
    \label{fig:suzakuregions}
\end{figure}

A similar approach was adopted for the \emph{Suzaku} spectra \citep[see][]{degrandi16}. In Fig. \ref{fig:suzakuregions} we show the \emph{XMM-Newton} image together with the definition of the regions used to study the spectral properties of the source using \emph{Suzaku} data. The region used to extract the background spectrum is shown as well. Note that the Eastern filament of A2744 \citep{eckert15} is located close to the radio relic; this region was masked during the analysis (see Fig. \ref{fig:suzakuregions}).

\begin{figure*}
\hbox{\includegraphics[width=\columnwidth]{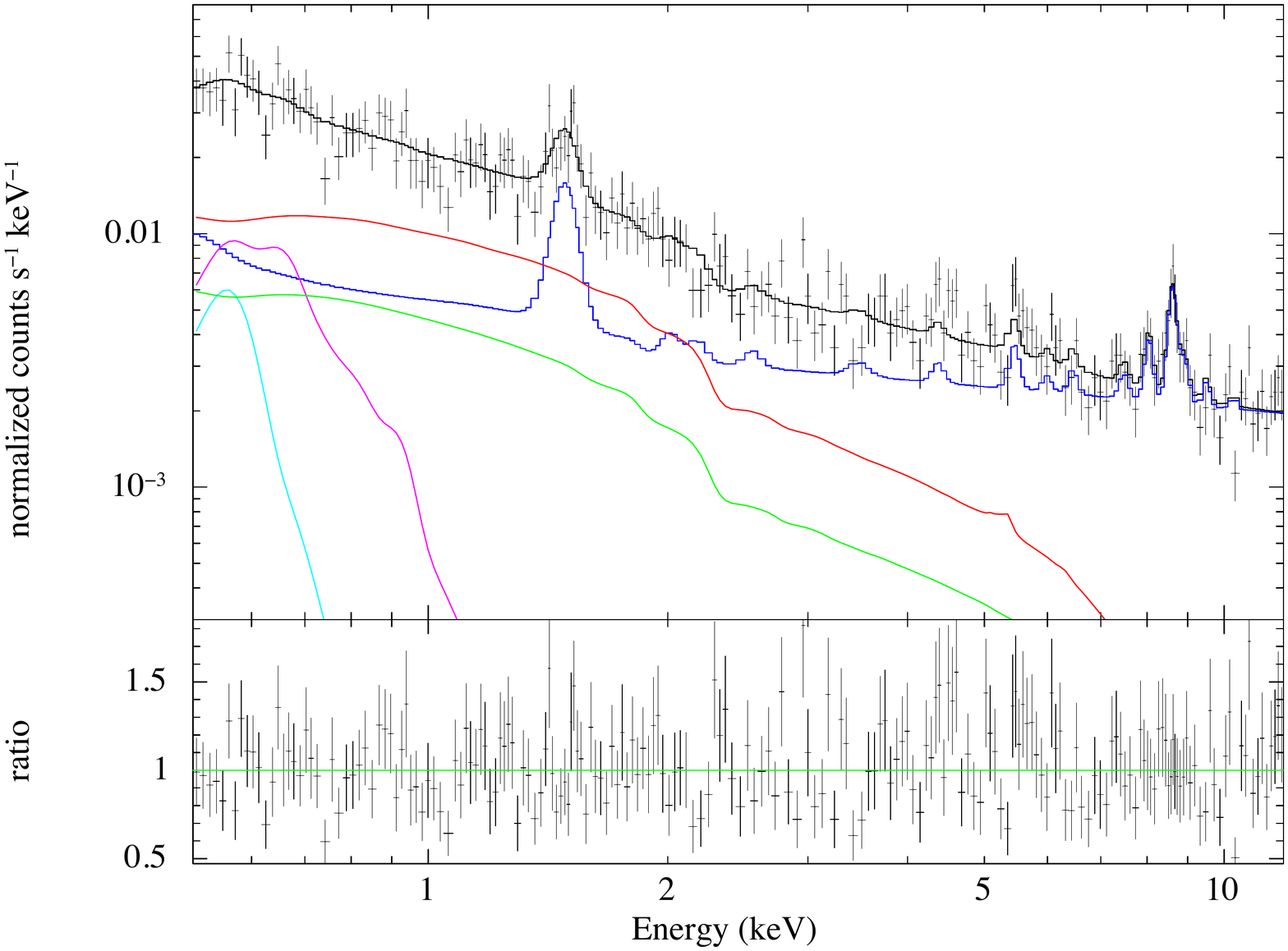}\includegraphics[width=\columnwidth]{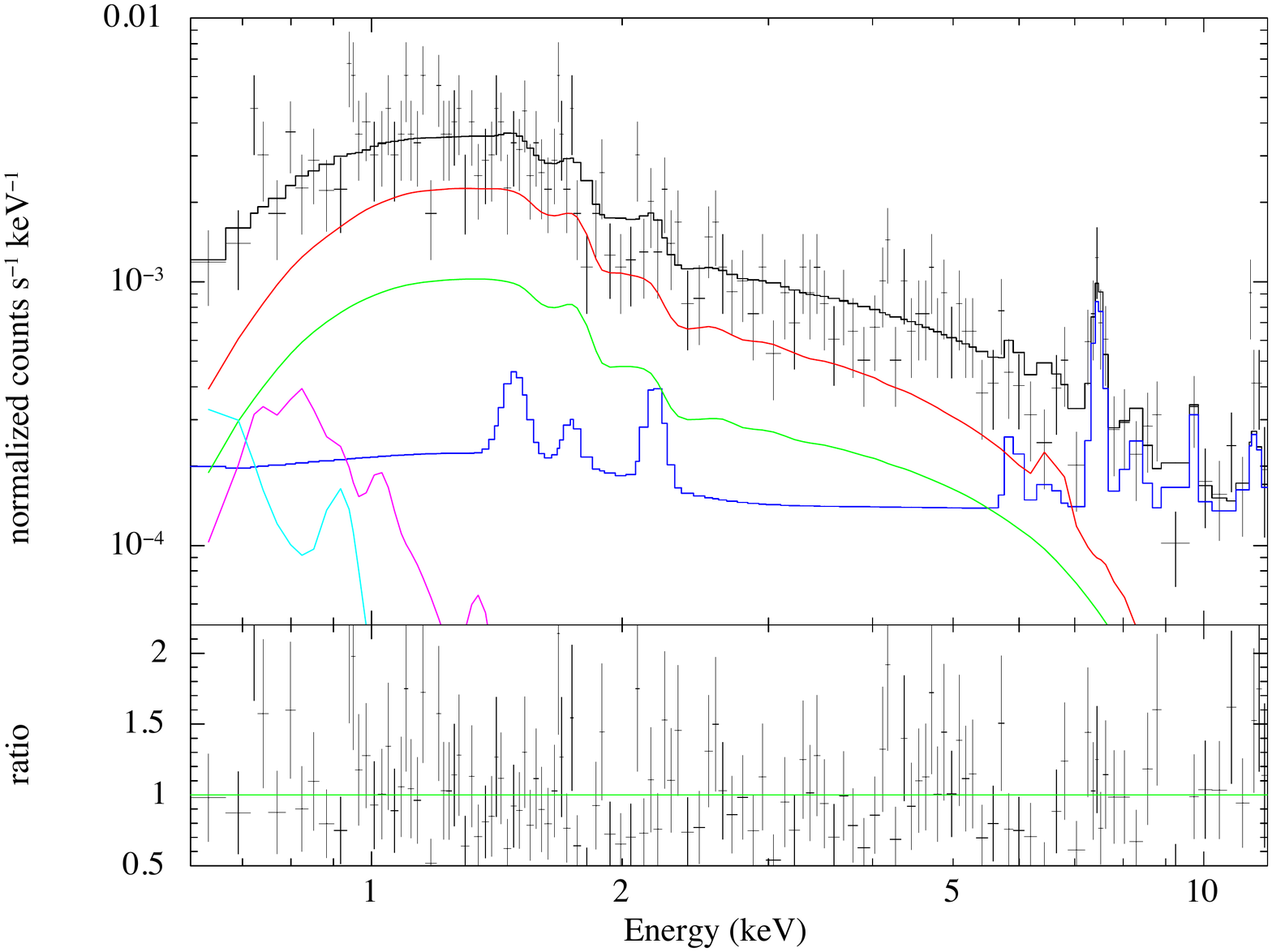}}
    \caption{X-ray spectra of the relic region obtained with \emph{XMM-Newton}/PN (left) and \emph{Suzaku}/XIS (right). The solid curves show the source (red), the cosmic X-ray background (green), the particle background (blue), the Galactic halo (magenta), the local hot bubble (cyan), and the total model (black). The bottom panels show the ratio between the data and the best-fit model. Only the \emph{XMM-Newton}/PN spectrum and the combined front-illuminated spectrum (XIS0+XIS3) are shown for clarity, however the fits were performed simultaneously on all available chips. }
    \label{fig:spec}
\end{figure*}

In Fig. \ref{fig:spec} we show the spectra of the relic region extracted with \emph{XMM-Newton} and \emph{Suzaku}, together with their best-fitting model. From the \emph{Suzaku} data we measured a projected temperature $kT=12.3_{-3.5}^{+4.5}$ keV. This value is consistent with the temperature reported by \citet{ibaraki14}. The fit to the \emph{XMM-Newton} data diverges to very high temperatures well outside the fitting range. The \emph{XMM-Newton} data allow us to set a lower limit of $kT>12.1$ keV to the temperature in the relic region (90\% confidence level). Therefore, the two instruments return consistent results and show that the temperature of the gas in the relic region is very high. For the remainder of the paper we adopt the \emph{Suzaku} value because of its lower particle background. 

\begin{figure}
\includegraphics[width=\columnwidth]{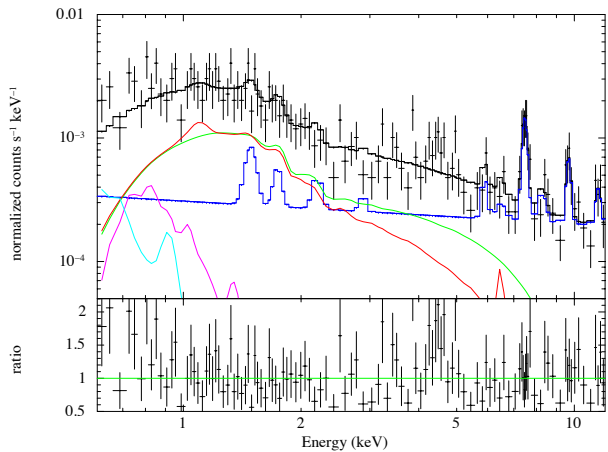}
    \caption{Same as Fig. \ref{fig:spec} for the \emph{Suzaku} spectrum of the pre-shock region.}
    \label{fig:preshock}
\end{figure}

We also extracted the \emph{Suzaku} spectrum from the pre-shock region (see Fig. \ref{fig:preshock}). Unfortunately, a number of relatively bright point sources are present in this area, which cover a large fraction of the region because of the broad PSF of \emph{Suzaku}. We fixed the position of these point sources to the position measured by \emph{XMM-Newton} and excluded circular regions of 1.5 arcmin radius around the corresponding positions. Fitting the spectrum of the remaining region following the same procedure as explained above, we obtained a projected temperature $kT=4.6_{-1.3}^{+2.3}$. Therefore, we conclude with high confidence that the density discontinuity shown in Fig. \ref{fig:jump} is a shock front with a projected temperature jump $T_{\rm in}/T_{\rm out}=2.7_{-0.9}^{+1.4}$.

An alternative explanation is that a fraction of the X-ray emission at the radio relic originates from inverse-Compton scattering of CMB photons with the relativistic electron population responsible for the radio relic. To test this possibility, we added an additional power-law component to model the \emph{Suzaku}/XIS spectrum. We fixed the spectral index of the model to the value of 1.1 reported by \citet{orru07} in the radio waveband and added the normalization of the power law as a free parameter. Adding this component yields a very modest improvement in C-statistic compared to the single APEC model ($\Delta$C=0.8) and the normalization of the power-law component is consistent with 0. We thus conclude that a strong non-thermal contribution to the X-ray emission is unlikely.

\section{Discussion}
\label{sec:disc}

\subsection{Mach number}

As shown in Fig. \ref{fig:jump}, we detect a surface-brightness edge at the position of the radio relic, which can be modeled with a projected broken power law with a density jump $r=n_{\rm in}/n_{\rm out}=1.9_{-0.4}^{+0.6}$. The density jump at the front can be related to the Mach number of the moving plasma through the Rankine-Hugoniot jump conditions \citep{markevitch07},

\begin{equation}\mathcal{M_{\rm shock}}=\left(\frac{2r}{\gamma+1-r(\gamma-1)}\right)^{1/2},\end{equation}

where $\gamma=5/3$ is the polytropic index. Using the above formula, we obtain a Mach number $\mathcal{M_{\rm shock}}=1.7_{-0.3}^{+0.5}$ at the front. Radio observations of the radio relic indicate a spectral index of $\alpha=1.1\pm0.1$ at the radio relic \citep{orru07}, which from the standard DSA formula, 

\begin{equation}\mathcal{M_{\rm radio}}=\left(\frac{p+2}{p-2}\right)^{1/2},\mbox{ with }p=2\alpha+1,\label{eq:injection}\end{equation}

translates into a Mach number $\mathcal{M_{\rm radio}}=2.1\pm0.1$, in agreement with the Mach number derived from the shock front, albeit slightly higher. Note however that this calculation assumes that all the electrons are freshly accelerated and neglects cooling losses. If we assume instead that the observed emission is dominated by older electrons previously accelerated by the same shock, the observed radio spectrum is related to the injection spectrum by $\alpha=\alpha_{\rm inj}+0.5$ \citep[][]{kardashev62}, which using the same formula gives a $\mathcal{M} \approx 4.5$ shock instead. In the following, we will also test this scenario, with the caveat that it is highly inconsistent with our X-ray estimate of the shock strength. Higher-resolution radio observations are required to test the consistency between the Mach number inferred from the X-ray shock and that expected from DSA theory.

The very high temperature measured in the relic region provides direct evidence of heating of the plasma induced by the passage of the shock wave. Although the relic is located at a rather large projected distance from the cluster core ($\sim1.5$ Mpc), the temperature of the relic region is higher than in the cluster core\footnote{We remind that A2744 does not host a cool core \citep[e.g.][]{owers11}, thus this statement is meaningful.}. Applying again the Rankine-Hugoniot jump conditions, one can predict the temperature jump at the front to lie in the range $1.4-2.1$, which is slightly lower (albeit consistent) with the measured projected temperature jump. Given the sound speed $c_s=\left(\gamma kT/\mu m_p\right)^{1/2}$ in the pre-shock plasma, we estimate the velocity of the shock wave to be in the range $2,000-2,800$ km/s.

We notice that the presence of substructures in the shock region might in principle affect both estimates of the Mach number, e.g. by introducing a clumping factor in temperature and density. However, both the observed large-scale regularity of the radio emission \citep[][]{orru07} and the theoretical expectations for the distribution of Mach numbers at the shock front under realistic cluster conditions \citep[][]{skillman13} suggest that this uncertainty is not large enough to reconcile X-ray and radio estimates of the Mach number, for the scenario in which the emission is dominated by aged electrons.

\subsection{Clues on electron acceleration efficiency}

The combination of X-ray and radio observations of A2744 allows us to test the DSA model for the origin of the radio relic. We follow the approach outlined in \citet{vazza14}, which makes it possible to constrain the acceleration efficiency of electrons at the shock to match the observed radio power. Given the relic parameters (i.e. the size and distance of the relic, the radio spectrum) and the gas density and temperature at the relic, we can estimate the kinetic energy flux across the shock and also how much energy should be dissipated into relativistic electrons to match the observed radio power. 

We fix the post-shock gas temperature to $12$ keV, while the pre-shock is derived from the shock jump conditions for the Mach numbers inferred by the X-ray ($\mathcal{M} \approx 1.7$) and the radio, in the extreme cases of fresh ($\mathcal{M} \approx 2.1$)  and old  ($\mathcal{M} \approx 4.5$) electrons. The gas density at the relic location is the extrapolation of the  $\beta$-model solution for A2744 at the relic position, and we add the density jump of the shock. For the magnetic field, we use the range of values derived by \citet{orru07} under the hypothesis of equipartition between cosmic ray electrons and magnetic fields,  $B \approx 0.6 - 1.1 \mu G$.  Table \ref{tab:sim} shows the list of values we investigated for this test. The last two columns show the electron acceleration efficiency, $\xi_e$  (defined as the fraction of kinetic energy flux across shocks which is dissipated into the acceleration of CR electrons), that we derive from the data and the corresponding value predicted by DSA. In the last case, we used the proton acceleration efficiency derived by the DSA model by \citet{kang13}, rescaled for an electron to proton injection ratio $K_{\rm ep}=0.01$, which is at the high end of what is assumed in DSA. The comparison clearly shows that the required electron acceleration efficiency for the weak shock in A2744 is from $10$ to $10^3$ times higher than  predicted by DSA, for the weak shocks where X-ray and radio estimates are consistent. In this case, the single injection model (i.e. Eq. \ref{eq:injection}) is inconsistent with our data.

The problem in explaining the radio emission from such weak shocks has been already addressed in the literature \citep{pinzke13,kang14,vazza14}, suggesting that a pool of mildly relativistic electrons must be already present in the volume, and that it can get reaccelerated by weak shocks. Given the variety of sources of relativistic electrons in the ICM (e.g. previous cluster shocks, supernovae, radio galaxies etc) this scenario is energetically viable. For example, the evidence of a physical connection between a peripheral radio relic and the lobes of a radio galaxy has been recently discussed by \citet{bonafede14}. The fact that the relic is so peripheral makes other alternative scenarios (i.e. a much higher magnetic field) very unlikely. However, the shock re-acceleration scenario also faces non-trivial problems, because unless the pre-existing electrons are the product of a leptonic-dominated acceleration mechanism ($K_{\rm ep} \geq 0.1$), shocks would also reaccelerate protons and cause $\gamma$-ray emission in excess of \emph{Fermi} limits \citep[e.g.][]{vazza15}.

Conversely, a shock with a Mach number $\mathcal{M} \approx 4.5$ does not represent a challenge for the injection of electrons since the predicted acceleration efficiency by DSA is large enough. However, in this case the Mach number inferred from the radio and the X-ray are inconsistent at the $\approx 6 \sigma$ level, posing a strong problem for the interpretation of these observations. 

\begin{table}
\label{tab:sim}
\caption{Adopted values for our modelling of shock electron acceleration.}
\centering \tabcolsep 5pt
\begin{tabular}{c|c|c|c|c|c|c}
$r$ & $n_{\rm pre}$ & $T_{\rm pre}$ & $M$ & $B_{\rm relic}$ & $\xi_e$ & $\xi_e$\\
 $\rm[Mpc]$ &   $\rm[1/cm^{-3}]$ & $[keV]$ &  & $\rm [\mu G]$ & (measured)&(predicted) \\ \hline
       1.55      &   $9.5 \cdot 10^{-5}$  & 7.0 & 1.7 & 1.1 & $4.6 \cdot 10^{-5}$ & $6.5 \cdot 10^{-8}$\\
       1.55      &   $9.5 \cdot 10^{-5}$  & 7.0 & 1.7 & 0.6 & $2.0 \cdot 10^{-4}$ & $6.5 \cdot 10^{-8}$\\
        1.55      &   $9.5 \cdot 10^{-5}$  & 5.4 & 2.1 & 1.1 & $3.8 \cdot 10^{-5}$ & $1.0 \cdot 10^{-7}$\\
         1.55      &   $9.5 \cdot 10^{-5}$  & 5.4 & 2.1 & 0.6 & $1.7 \cdot 10^{-4}$ & $1.0 \cdot 10^{-7}$\\
	1.55      &   $9.5 \cdot 10^{-5}$  & 1.2 & 4.5 & 1.1 & $4.3 \cdot 10^{-5}$ & $6.7 \cdot 10^{-4}$\\
         1.55      &   $9.5 \cdot 10^{-5}$  & 1.2 & 4.5 & 0.6 & $3.8 \cdot 10^{-5}$ & $6.7 \cdot 10^{-4}$\\
 \end{tabular}
\end{table}

\section{Conclusions}

In this paper, we have reported the discovery of a density jump by a factor of $\sim1.9$ associated with the eastern edge of the radio relic of A2744. The temperature of the gas coincident with the radio relic is very high ($\sim12$ keV), which indicates a heating of the plasma by the passage of a shock front. Our data imply a Mach number of $1.7_{-0.3}^{+0.5}$ at the front. This value agrees with the temperature jump and with the Mach number of $2.1$ expected from the spectral index of the radio relic \citep{orru07}, although this estimate assumes that all electrons are freshly accelerated. The high temperature of the plasma in the post-shock region provides clear evidence for simultaneous heating and particle acceleration by merger-induced shock waves. The acceleration efficiency implied by DSA is insufficient to explain the observed radio power, which suggests that pre-existing mildly-relativistic electrons have been re-accelerated at the shock front. Finally, we note that the shock front reported here should be an excellent target for the new generation of high-resolution Sunyaev-Zeldovich cameras (e.g. NIKA2), which could allow us to detect the associated pressure jump directly.

\section*{Acknowledgements}

Based on observations obtained with \emph{XMM-Newton}, an ESA science mission with instruments and contributions directly funded by ESA Member States and NASA. GMRT is run by the National Centre for Radio Astrophysics of the Tata Institute of Fundamental Research. MJ was supported by the Science and Technology Facilities Council [grant number ST/L00075X/1 \& ST/F001166/1]. FV acknowledges support from the grant VA 876/3-1 and FOR1254 from the Deutsche Forschungsgemeinschaft. MSO acknowledges the funding support of the Australian Research Council through a Future Fellowship (FT140100255). JPK acknowledges support from the ERC advanced grant LIDA. 

\vspace{-0.2cm}
\bibliographystyle{mnras}
\bibliography{a2744_relic}



\bsp	
\label{lastpage}
\end{document}